\documentstyle[12pt,preprint,tighten,aps]{revtex}
\newtheorem{defi}{Definition}

\def\ut#1{\rlap{\lower1ex\hbox{$\sim$}}{#1}}
\def\contr#1{\raise1ex\hbox{\small $[#1]$}}
\def\H{{\cal H}}
\def\SO{{\rm SO}}
\def\R{{I\!\!R}}
\def\C{{\bf C}}

\preprint{\vbox{\baselineskip=12pt
\rightline{\vbox{\hbox{CGPG-99/3-2}
\hbox{NSF-ITP/99/19}}}}}

\begin{document}

\title{Simple Spin Networks as Feynman Graphs}
\author {L.\ Freidel${}^{1,2}$, K.\ Krasnov${}^{1}$\thanks{%
E-mail addresses: freidel,krasnov@phys.psu.edu}}
\address{1. Center for Gravitational Physics and Geometry \\
Department of Physics, The Pennsylvania State University \\
University Park, PA 16802, USA}

\address{2. Laboratoire de Physique  \\
Ecole Normale Sup\'erieure de Lyon \\
46, all\'ee d'Italie, 69364 Lyon Cedex 07, France}
\date{\today}
\maketitle

\begin{abstract}

We show how spin networks can be described and evaluated
as Feynman integrals over an internal space.
This description can, in particular, be applied to the
so-called simple $\SO(D)$ spin networks that are of importance for
higher-dimensional generalizations of loop quantum gravity.
As an illustration of the power of the new formalism, we use it
to obtain the asymptotics of an amplitude for the $D$-simplex
and show that its oscillatory part is given by the Regge action.

\end{abstract}

\section{Introduction}
\label{sec:intr}

Spin networks were originally introduced by Penrose \cite{Penrose}
in an attempt to give a combinatorial description of spacetime. Since
then they reappeared in many branches of mathematical physics, see,
e.g., \cite{future} and references therein. In particular, spin
networks are of fundamental importance in the loop approach to
quantum gravity \cite{Loop,Roberto}. More recently, with a development of
the formalism of spin foam quantization \cite{Mike,BC,Baez,SF}, it
was realized that spin networks that appear in the path integral
version of loop quantum
gravity are of a very special type. In the case of four spacetime
dimensions these special spin networks were discovered in
\cite{BC,Baez}. Their higher-dimensional analogs were then
described in \cite{FKP}. These spin networks can be called
{\it simple}:
they satisfy a quantum analog of the  simplicity constraint
which requires a bivector to be a wedge product and,
as was realized in \cite{FKP}, these are in
certain precise sense the simplest possible spin networks that can
be constructed in a given dimension.

In this paper we give a
new description of simple spin networks: we show that
they can be viewed and evaluated as Feynman graphs. Our construction gives an
interesting perspective on these objects, which are of fundamental
importance in the loop approach to quantum gravity. It also gives
us new technical tools: with the help of the Feynman graph
description we will be able to obtain the asymptotics of
an amplitude for a $D$-simplex. We find the expected
Regge action asymptotics for the amplitude.

The organization of
this paper is as follows. In the rest of this section we remind
the reader the usual description of spin networks
and give the main idea of our construction. It turns out that
the description of spin networks as Feynman graphs is
quite general and relevant to gauge groups other than $\SO(D)$
considered in \cite{FKP}. The main idea of our construction
is quite simple even when presented in its full generality.
Thus, in this section, we do not restrict ourselves to any
particular choice of the gauge group and present a general
construction. It is then applied in Sec. \ref{sec:simple} to simple $\SO(D)$
spin networks relevant for  quantum gravity. As an illustration of
the usefulness of the new description we utilize it to derive
the asymptotics of the evaluation of a simple $\SO(D)$ spin network in
Sec. \ref{sec:asympt}.

Let us now remind the reader the standard description of spin
networks, see, e.g., \cite{BaezSpinNet}.
\begin{defi} Given a Lie group
$G$ (which we assume to be semisimple and compact),
spin network is a triple $(\Gamma,\rho,I)$, where:\\
\noindent(i) $\Gamma$ is an oriented graph;\\
\noindent(ii) $\rho$ is a labelling of
each edge $e$ by an irreducible unitary representation $\rho_e$
of $G$;\\
\noindent(iii) $I$
is a labelling of each vertex $v$ of $\Gamma$
by an intertwiner $I_v$ mapping the
tensor product of incoming representations at $v$ to the product
of outgoing representations at $v$.
\end{defi}
\noindent These data define a function
$\phi_{(\Gamma,\rho,I)}$ on $G^E$  invariant under the
action of the group at vertices; here $G^E$ is the product of a number
of copies of the group $G$, one for each of $E$ edges of $\Gamma$.
Thus, spin network can be described as a function which associates a number
to each
assignment of group elements $g_e$ to the edges $e$:
\[
\phi_{(\Gamma,\rho,I)}: G^E \to \C
\]
An explicit
construction of this function proceeds as follows. First, for
each edge $e$ let us consider the operator representing
the group element $g_e$ in the representation $\rho_e$.
Introducing a basis in the corresponding representation space
one can calculate the matrix elements $U^{\rho_e}(g_e)_m^n$.
One then takes a tensor product of all these matrix elements
to obtain a tensor $U(g^E)$; it has one subscript and
one superscript for each edge $e$. Then for each vertex $v$
of $\Gamma$ let $S(v)$ be the set of edges having $v$
as ``source'' and let $T(v)$ be the set of edges having
$v$ as ``target''. The intertwiner is a map
(commuting with the action of $G$):
\[
I_v: \otimes_{e\in T(v)} \rho_e \to \otimes_{e\in S(v)} \rho_e.
\]
We can think of $I_v$ as of a tensor with one superscript
for each edge $e\in T(v)$ and one subscript for each edge
$e\in S(v)$. One can then form the tensor product of all
intertwiners to obtain a tensor $I$, and then take the
tensor product $U(g^E)\otimes I$. Note now that each superscript
in $U$ corresponds to a subscript in $I$ and vice versa, because
each edge of $\Gamma$ lies in $S(v)$ for one vertex $v$ and in
$T(w)$ for one vertex $w$. Therefore, one can contract indices
of $U(g^E)\otimes I$ to get a number. This is the
value of the function $\phi_{(\Gamma,\rho,I)}$. One can
check that the function constructed is invariant under the
action of the group $G$, where the group action is that at
the vertices.

Before we present our construction we will need
the standard notion of a {\it representation of class 1}
(see, e.g., \cite{VK}). Spin networks that can be represented
as Feynman graphs are the ones constructed using only these special
representations of $G$.
\begin{defi} Let $\rho$ be an irreducible representation of
$G$, and let $H$ be a subgroup of $G$. If the representation
space $V^\rho$ contains vectors invariant under $H$, and if all
operators $U^\rho(h), h\in H$ are unitary, then $\rho$ is
called a {\it representation of class 1} with respect to $H$.
\end{defi}
\noindent The significance of these representations comes from
the fact that they can be realized in the space of functions
on the homogeneous space $H\backslash G$. As we describe below,
spin networks that are constructed using only representations
of class 1 with respect to $H$ can be viewed as Feynman graphs
on $H\backslash G$. Simple $\SO(D)$ spin networks of \cite{FKP}
are just such spin networks. In this case $H=\SO(D-1)$ and
$H\backslash G = S^{D-1}$.

The realization of a representation of class 1 in the space of
functions on a homogeneous space $H\backslash G$ is a particular
case of a general description of an irreducible representation $\rho$
by shift operators in the space of functions on the group.
Let us remind the reader this description. Consider matrix elements
\[
U^\rho_{\bf x,a}(g) := (U^\rho(g){\bf x}, {\bf a}),
\]
where $\bf x,a$ are vectors from the representation space $V^\rho$.
Let us fix $\bf a$. Then the functions $U^\rho_{\bf x,a}(g), {\bf x}\in V^\rho$
span a subspace in the space $L^2(G)$ of square integrable functions on the
group. One can then show that the right regular action of the
group $G$ on this subspace gives an irreducible representation
equivalent to $\rho$. The scalar product in the representation
space is then given by the integral over the group.
In the case $\rho$ is a representation
of class 1 with respect to $H$, and $a$ is a vector invariant
under $H$, the functions $U^\rho_{\bf x,a}(g)$ are constant on
the right cosets $H g$  and can be regarded as functions on
the homogeneous space $X=H\backslash G$. The scalar product is then
given by an integral over $X$.

We are now ready to describe spin networks constructed from
representations of class 1 with respect to $H$ as Feynman graphs
on $X$. Let us denote by $P^{(\rho) n}(x), x\in X$
an orthonormal basis in the  representation space $\rho$ realized in the
space of functions on $X$. The matrix elements of
the group operators are then given by:
\[
U^\rho(g)_m^n = \int_X dx \, \overline{P_m^{(\rho)}(x)}
P^{(\rho) n}(x g),
\]
where $dx$ is the invariant normalized measure on $X$. This gives
realization of the matrix elements as integrals over $X$.
The other building block necessary to construct a spin network
is an intertwiner. Intertwiners can be characterized by their
integral kernels. For a $k$-valent vertex one defines
the integral kernel $I_v(x_1,\ldots,x_k)$ so that:
\begin{eqnarray}\nonumber
&{}&I_{v\, m_1\ldots m_i}^{\,\,\,n_{i+1}\ldots n_k} = \\\nonumber
&{}&\int_X dx_1 \cdots dx_k \,
I_v(x_1,\ldots,x_k) \,
\overline{P_{m_1}^{(\rho_1)}(x_1)} \cdots
\overline{P_{m_i}^{(\rho_i)}(x_i)} P^{(\rho_{i+1})n_{i+1}}(x_{i+1}) \cdots
P^{(\rho_k) n_k}(x_k).
\end{eqnarray}
The integral kernels $I_v(x_1,\ldots,x_k)$ must satisfy the invariance
property $I_v(x_1 g,\ldots,x_k g) = I_v(x_1,\ldots,x_k)$.
A special important set of intertwiners is given by:
\[
\tilde{I}_v(x_1,\ldots,x_k)=\int_X dx \, \delta(x,x_1)\cdots\delta(x,x_k)
\]
or
\[
\tilde{I}_{v\, m_1\ldots m_i}^{\,\,\,n_{i+1}\ldots n_k} =
\int_X dx \,
\overline{P_{m_1}^{(\rho_1)}(x)} \cdots
\overline{P_{m_i}^{(\rho_i)}(x)} P^{(\rho_{i+1})n_{i+1}}(x) \cdots
P^{(\rho_k) n_k}(x).
\]
These special intertwiners are the ones that appear in simple
spin networks of \cite{FKP}. Exactly for such intertwiners it
is possible to represent the spin network evaluation as a
Feynman graph. Let us now introduce what can be called Green's function:
\[
G^{(\rho)}(x,y) := \sum_n \overline{P_n^{(\rho)}(x)} P^{(\rho) n}(y).
\]
This Green's function satisfies the ``propagator'' property:
\[
\int_{X} dz \, G^{(\rho)}(x,z) G^{(\rho)}(z,y) = G^{(\rho)}(x,y).
\]
Let us also introduce a propagator ``in the presence of
a source'':
\[
G^{(\rho)}(x,y; g): = \int_X dz \, G^{(\rho)}(x,z)
G^{(\rho)}(z g, y).
\]
It is clear that $G^{(\rho)}(x,y; e) = G^{(\rho)}(x,y)$,
where $e$ is the identity element of the group. One can now check that,
in the case all spin network intertwiners are of a
special type $\tilde{I}_v$ described above,
the spin network function $\phi_{(\Gamma,\rho,\tilde{I})}$ of
the group elements $g_1,\ldots,g_E$ is given by the
Feynman graph with the following set of Feynman rules:
\begin{itemize}
\item With every edge $e$ of the graph $\Gamma$ associate a
propagator $G^{(\rho_e)}(x,x'; g_e)$.
\item Take a product of all these data and integrate over one copy of $X$
for each   vertex.
\end{itemize}
These rules can be summarized by the following formula:
\begin{equation}\label{**}
\phi_{(\Gamma,\rho,\tilde{I})} (g_1,\ldots,g_E) =
\prod_v \int_X dx_v
\prod_e G^{(\rho_e)}(x,x'; g_e).
\end{equation}
Thus, in the case intertwiners are given by $\tilde{I}$
the evaluation of a spin network on a string of group
elements is given by a Feynman graph: one associates
the Green's function to every edge and integrates over the
positions of vertices.

Before we illustrate this general construction on the example
of simple $\SO(D)$ spin networks, let us note
that this construction can be readily generalized to the
case of an arbitrary spin network. Indeed, the restriction
of representations labelling the spin network to be those
of class 1 with respect to a fixed subgroup $H$ was necessary
only to guarantee that the resulting Feynman graph lives in
the homogeneous space $X=H\backslash G$. It can be dropped
at the expense of Feynman graphs becoming graphs in the
group manifold. The restriction of intertwiners to be of
a special type $\tilde{I}_v$ can be dropped with the result
that the set of Feynman rules specified above changes:
in this case one has to associate with every vertex the
integral kernel $I_v(x_1,\ldots,x_k)$ and then integrate
over all the arguments. Thus, in the case of arbitrary
intertwiners, the evaluation formula takes the form:
\begin{equation}\label{*}
\phi_{(\Gamma,\rho,I)} (g_1,\ldots,g_E) =
\prod_v \int_X d{\bf x}_v \, I_v({\bf x}_v)
\prod_e G^{(\rho_e)}(x,x'; g_e).
\end{equation}
Here ${\bf x}_v$ stands for a string of arguments $x_1,\ldots,x_k$
of a $k$-valent intertwiner, and $x,x'$ in the argument of the
Green's function $G^{(\rho_e)}(x,x'; g_e)$ must be the same as
those in two intertwiners: $x$ must the appropriate argument
in $I_v, e\in S(v)$ and $x'$ must be the argument of $I(w), e\in T(w)$.

\section{Simple $\SO(D)$ spin networks}
\label{sec:simple}

In this section we illustrate the general construction
presented above on the example of simple $\SO(D)$
spin networks. Their relevance to quantum gravity in
$D$ dimensions was explained in \cite{FKP}.

Simple $\SO(D)$ spin networks are the ones constructed from
special representations of $\SO(D)$. As is well-known, group
$\SO(D)$ has a special class of representations,
called  spherical harmonics, that appear in the decomposition of
the space of functions  $L^2(S^{D-1})$ on  $S^{D-1}$
into irreducible components.
Some properties of these representations
are described in Appendix \ref{app:simple}. Using the terminology
introduced in
Sec. \ref{sec:intr} these representations of $\SO(D)$ can be
described as representations of class 1 with respect to
$\SO(D-1)$.
They are characterized by a single parameter that we will
denote by $N$ in what follows; $N$ is required to be an
integer. These are the representations that were called
simple in \cite{FKP}. A simple $\SO(D)$ spin network was
defined in \cite{FKP} as a spin network which is constructed
only from simple representations and whose intertwiners are
the special intertwiners $\tilde{I}$ introduced in Sec. \ref{sec:intr}.

In the case intertwiners are given by $\tilde{I}$,
the value of a simple spin network on a sequence of group
elements can be evaluated using the general formula (\ref{**}).
In what follows we will
be concerned only with a special case of spin network
evaluated on all group elements being equal to the identity
element. This ``evaluation'' of a spin network gives
a number that depends only on the graph and on the
labelling of its edges by integers $N_e$. Evaluation
of a spin network is of special importance for
quantum gravity because this is the way to obtain
an amplitude for a spacetime simplex, see \cite{BC,Baez,SF}.
Thus, according to our Feynman graph formula (\ref{**}),
the evaluation of a simple spin network is given by
\begin{equation}\label{evaluation}
\phi_{(\Gamma,\rho)} = \prod_v \int_{S^{D-1}} dx_v
\prod_e G_{N_e} (x,x').
\end{equation}
Here
\begin{equation} \label{kern}
G_N (x,y) =\sum_{K} \overline{\chi_{K}(x)} \chi^{K}(y),
\end{equation}
where we have introduced an orthonormal basis
$\chi^{K}$, $ K=(k_1,\cdots, k_{D-2})$ $N\geq k_1
\cdots \geq k_{D-3} \geq |k_{D-2}| $ in the representation
space (see Appendix \ref{app:simple} for a construction
of such a basis). The invariance property
$G_N(x g, y g)=G_N(x,y)$ implies that $G_N(x,y)$ depends only on the
scalar product $(x\cdot y)$, and it is a standard result \cite{VK}
that
\begin{equation}\label{Gege}
G_N^{(D)}(x,y) = { D+2N-2 \over D-2} C_{N}^{(D-2)/2}(x\cdot y),
\end{equation}
where $ C_{N}^{p}$ is the Gegenbauer polynomial, see Appendix
\ref{app:geg} for the definition. The expression (\ref{evaluation})
for the evaluation of a simple spin network is a generalization
of the result \cite{Barrett:evaluation} for the evaluation in
$D=4$.

\bigskip
\noindent{\bf Example: Evaluation of the $\Theta$-graph}
\bigskip

Let us use the above representation of the simple spin network
evaluation to compute the evaluation of the $\Theta$-graph.
This is of importance because the value of the $\Theta$-graph
appears in the normalization of tri-valent vertices. According
to (\ref{evaluation}) the evaluation is given by:
\begin{equation}\label{Tint}
\Theta^{(D)}(N_1,N_2,N_3)= \int dx dy \,
 G_{N_1}(x,y) G_{N_2}(x,y) G_{N_3}(x,y).
\end{equation}
Using the expression of $G_N$ in terms of a Gegenbauer polynomial
(\ref{Gege}), this integral can be computed. It is not
equal to zero only if $g=(N_1+N_2+N_3)/2$ is  an integer and
$g-N_i\geq 0, i=1,2,3$. In this case one gets:
\begin{equation}\label{TGam}
\Theta^{(D)}(N_1,N_2,N_3) =
{\Gamma(g+2p) \Gamma(p+1) \over \Gamma(g+p+1) \Gamma(2p)}
\prod_{i=1}^3 \left(
{(N_i+p) \Gamma(g-N_i+p) \over \Gamma(p+1) \Gamma(g-N_i+1)}
\right).
\end{equation}
Here $p =(D-2)/2$. To check this result one can check that (\ref{Tint}),
(\ref{TGam}) both satisfy the recurrence relation (implied by
(\ref{rec})):
\begin{eqnarray}
	{N_1 +1 \over N_1 +p +1 } \Theta^{(D)}(N_1+1,N_2,N_3) +
	{N_1 +2p-1 \over N_1 +p -1} \Theta^{(D)}(N_1-1,N_2,N_3) = \\
	{N_3 +1 \over N_3 +p +1} \Theta^{(D)}(N_1,N_2,N_3+1) +
	{N_3 +2p-1 \over N_3 +p -1} \Theta^{(D)}(N_1,N_2,N_3-1),
\end{eqnarray}
and that $\Theta(N_1,N_2,0)$ reproduces the
orthogonality relation (\ref{norm}).
For $D=4$ the above expression simplifies:
\begin{equation}\label{theta4}
\Theta^{(4)}(N_1,N_2,N_3) = (N_1+1)(N_2+1)(N_3+1).
\end{equation}

This result can be used to show that the intertwiner used
in \cite{FKP} to define simple spin networks in the case
of $D=4$ coincides with the one proposed in \cite{BC}. Indeed,
the four-valent intertwiner of \cite{FKP} reads:
\[
I_{2,2}(\overline{P_1},\overline{P_2}, Q_1, Q_2) =
\int dx\, 
\overline{P_1(x)} \overline{P_2(x)} Q_1(x) Q_2(x).
\]
Using the kernel of the identity operator on $L^2(S^{D-1})$
given by $\sum_{N=0}^\infty G_N$ we can expand the 4-valent vertex in
terms of the sum of the product of two tri-valent ones:
\[
I_{2,2}(\overline{P_1},\overline{P_2},Q_1,Q_2) =
\sum_{N=0}^{\infty} \int dx dy \,
\overline{P_1(x)}\overline{P_2(x)} G_N(x,y) Q_1(y) Q_2(y).
\]
In short this can be written as
\begin{eqnarray}\nonumber
I_{2,2}^{(D)}(N_1,N_2,N_3,N_4) =
\sum_{N} I_{2,1}^{(D)}(N_1,N_2,N) \cdot I_{1,2}^{(D)}(N,N_3,N_4) = \\ \nonumber
\sum_{N} \left[ \Theta^{(D)}(N_1,N_2,N) \Theta^{(D)}(N,N_3,N_4) \right]^{1/2}
\bar{I}_{2,1}^{(D)}(N_1,N_2,N) \cdot \bar{I}_{1,2}^{(D)}(N,N_3,N_4),
\end{eqnarray}
where we have introduced the  normalized intertwiner
\[
{\bar I}_{2,1}^{(D)}(N_1,N_2,N_3) =
(\Theta^{(D)}(N_1,N_2,N_3))^{-1/2}{ I}_{2,1}^{(D)}(N_1,N_2,N_3).
\]

In the case of $D=4$, using the result (\ref{theta4}) for the
$\Theta$-graph, we get
\begin{eqnarray}\nonumber
&{}& I_{4}^{(D)}(N_1,N_2,N_3,N_4) = \\ \nonumber
&{}& \left[ (N_1+1) (N_2+1) (N_3+1) (N_4+1)  \right]^{1/2} \sum_{N=0}^\infty
(N+1) \bar{I}_{2,1}^{(D)}(N_1,N_2,N) \cdot \bar{I}_{1,2}^{(D)}(N,N_3,N_4).
\end{eqnarray}
Up to an overall normalization factor, this vertex is exactly
the vertex given in \cite{BC}.

\section{Large spin asymptotics}
\label{sec:asympt}

In this section we use the Feynman graph representation of the
simple spin networks to study the asymptotics of a $D$-simplex
amplitude for large $N$. The results of this section generalize
those of \cite{Barrett:asymptotics} to the case of arbitrary
dimension. Most of the labor necessary to get the asymptotics
is done in Appendix \ref{app:asympt}. Here we simple use the
asymptotics (\ref{app:as}) of the Gegenbauer polynomial obtained there.

As is explained in Refs. \cite{FKP}, the amplitude for a $D$-simplex
is given by the evaluation of the spin network that is dual to
the boundary of the simplex. The $(D-2)$-simplices are labelled
by simple representations of $\SO(D)$, i.e., by integers $N$. The
edges of the spin network dual to the boundary of the simplex
are in one-to-one correspondence with the $(D-2)$-simplices,
and inherit the labels of $(D-2)$-simplices. As one can easily
check, all vertices of the spin network in question are $D$-valent.
All intertwiners are of the special type described in Sec. \ref{sec:intr},
and, thus, the formula (\ref{evaluation}) can be used for
the evaluation. Using the asymptotics (\ref{app:as}) and
the formula (\ref{evaluation})  we present an asymptotic  evaluation
the amplitude: we will use the stationary phase
approximation for the integral. Our discussion follows
closely that of \cite{Barrett:asymptotics}.

To get a feeling about the behavior of the amplitude, we
will concentrate only on the oscillatory part of $C_N^p(\cos\theta)$.
Thus, dropping all multiplicative constants, which are unimportant for us,
we get
\[
\phi_{(\Gamma,\rho)} \sim \sum_{\{\epsilon_{kl}\}}
\left( \prod_{k<l} \epsilon_{kl} \right)
\int_{S^{D-1}} dx_1 \cdots dx_{D+1} \,
e^{i\sum_{k<l} \epsilon_{kl} ((N_{kl}+p)\theta_{kl}+(1-p)\pi/2)},
\]
where the integral is taken over $(D+1)$ points -- vertices of the
spin network -- on the unit $(D-1)$-sphere, and $k,l$ are indices
labelling the vertices $k,l=1,\ldots,D+1$. Thus, a pair $kl$ labels
a spin network edge, and $\theta_{kl}: \cos\theta_{kl}=x_k\cdot x_l$.
The quantity $\epsilon_{kl}$ takes values $\pm1$ and the sum is
taken over both possibilities for every edge. The rest of the
analysis is exactly the same as in \cite{Barrett:asymptotics}.
Taking into account the fact that the variation of the angles
satisfy the following  identity (see \cite{Barrett:asymptotics}):
\[
	\sum_{k<l} V_{kl} \delta \theta_{kl} =0,
\]
where $V_{kl}$ are the volumes of $(D-2)$-simplices inside
a geometric $D$-simplex, one finds that all $\epsilon_{kl}$ are either positive or
negative, and that the stationary phase values of $\theta_{kl}$
are the ones corresponding to a geometric $D$-simplex determined
by $N_{kl}+p $ interpreted as volumes of $(D-2)$-simplices. Then,
in the case the number $D(D+1)/2$ of edges in the simplex is even, we get
\begin{equation}\label{ass:1}
\phi_{(\Gamma,\rho)} \sim
\cos\left( \sum_{k<l} (N_{kl}+p) \theta_{kl} + \kappa {\pi\over 4} \right),
\end{equation}
where $\theta_{kl}$ are the higher-dimensional analogs
of the dihedral angles of the geometric $D$-simplex determined
by $N_{kl}+p$ and
\[
\kappa={(D+1)D\over 2}\,(4-D)
\] 
is the integer determined by $D$. In the case $D(D+1)/2$ is odd
one gets `$\sin$' instead of `$\cos$' in the asymptotics (\ref{ass:1}).
Thus, the simplex amplitude has the asymptotics of the exponential of the
Regge action, as expected.

\section{Acknowledgements}

We would like to thank Raymond Puzio for the help on the 
initial stage of this work, and Sameer Gupta for reading the
manuscript. This work was supported in part by the NSF
grant PHY95-14240 and by the Eberly research funds of Penn State.
K.K. was supported in part by a Braddock fellowship from Penn State.
We are also grateful to the Institute for Theoretical Physics,
University of California, Santa Barbara, where this work was
done. We acknowledge the support from NSF grant No. PHY94-07194.

\appendix
\section{Simple representations of $\SO(D)$}
\label{app:simple}

What is referred to in this paper as simple representations
of $\SO(D)$ are the usual spherical harmonics representations.
They are irreducible representations of $\SO(D)$ of class 1 with
respect to the subgroup $\SO(D-1)$ and, therefore, can be realized
in the space of functions on $S^{D-1}$. This partially explains
their relevance for quantum gravity in $D$ dimensions, where
the $(D-1)$-sphere has the geometrical meaning of the boundary
of the $D$-simplex. In this Appendix we review some basic
properties of these representations. For more information
see, e.g., \cite{VK}.

The spherical harmonics representations of $\SO(D)$ are the
most obvious ones: they can be realized in the space of
homogeneous polynomials of degree $N$. Let us denote the
space of such polynomials by $V_N^{(D)}$. Then
\[
{\rm dim} V_N^{(D)} = {(N+D-1)!\over N! (D-1)!}.
\]
It turns out, however, that the representation in this space
is not irreducible. The invariant subspace in $V_N^{(D)}$
is given, as usual, by the space of polynomials satisfying
the Laplace equation in $\R^D$. Thus, the irreducible representations
of this type are realized in the space of homogeneous harmonic
polynomials of degree $N$. Let us denote this space by
$\H_N^{(D)}$. As one can show,
\begin{equation}\label{app:dim}
{\mathrm dim}{\H^{(D)}_N} = {(2N+D-2)(D+N-3)!\over(D-2)!N!}.
\end{equation}

As we have mentioned, these representations are of the class
1 with respect to $\SO(D-1)$. Choosing the upper-left corner
embedding of $\SO(D-1)$ into $\SO(D)$, the vector in $\H_N^{(D)}$
that is invariant under the action of $\SO(D-1)$ is given
(up to normalization) by $C_N^p(x_D)$ for $x=(x_1,\ldots,x_D)$.
Here $p=(D-2)/2$ and $C_N^p(x)$ is the so-called Gegenbauer
polynomial defined in the next Appendix.

An explicit basis in $\H_N^{(D)}$ can be constructed by
choosing a string of embeddings
\[
\SO(2)\subset\SO(3)\subset\cdots\SO(D-1)\subset\SO(D).
\]
Then $H_N^{(D)}$ decomposes into subspaces irreducible
with respect to the action of the subgroup $\SO(D-1)$. The later
again decompose into the irreducible subspaces with
respect to the action of $\SO(D-2)$ etc. Finally, one
arrives at $\SO(2)$ whose irreducible representations
are 1-dimensional. Thus, we have:
\[
\H_N^{(D)} = \oplus_{k_1=0}^{N}\oplus_{k_2=0}^{k_1}
\cdots\oplus_{k_{D-2}=-k_{D-3}}^{k_{D-3}} V_{k_{D-2}}.
\]
Here $V_k$ are 1-dimensional representation spaces of $\SO(2)$.
Note that $k_{D-2}$ in the last sum runs over both positive
and negative values. Thus, a basis in $\H_N^{(D)}$ can
be labelled by a string of integers:
\[
K:=(k_1,k_2,\ldots,k_{D-2}),\qquad N\geq k_1 \geq k_2 \geq \cdots \geq
|k_{D-2}|.
\]

\section{Properties of Gegenbauer polynomials}
\label{app:geg}

Gegenbauer polynomials are orthogonal polynomials
satisfying many different properties. In this Appendix
we review some of them. For more information of Gegenbauer
polynomials see, e.g., \cite{VK,Hos}.

Let $p$ be denote a quantity related to the dimension $D$
according to $p=(D-2)/2$, or $D=2p+2$.
A  generating functional  for Gegenbauer polynomial is given by:
\begin{equation}\label{geg:gen}
(1-2xr+r^2)^{-p}= \sum_{N=0}^{+\infty} C_{N}^{p}(x) r^N.
\end{equation}
Gegenbauer polynomials satisfy the Rodriguez formula:
\begin{eqnarray}
C_{N}^{p}(x)= {(-1)^N (N+2p-1)(N+2p-2)\cdots (2p) \over
2^N N! (N+p-{1\over2}) (N+p-{3\over2})\cdots (p+{1\over 2})}
\times \\ \nonumber
(1-x^2)^{-p+{1\over 2}} ({d\over dx})^{N} (1-x^2)^{N+p-{1\over 2}},
\end{eqnarray}
where the prefactor can also be written as
\[
{(-1)^N \over 2^N N!} {\Gamma(N+2p) \Gamma(p+{1\over2}) \over
\Gamma(2p) \Gamma(N+{1\over2}+ p) }
\]
The recurrence formula is given by:
\begin{equation}\label{rec}
      (N+1)  C_{N+1}^{p}(x)- 2(N+p) x  C_{N}^{p}(x)
        + (N+2p -1) C_{N-1}^{p}(x) =0,
\end{equation}
with $C_{0}^{p}(x) =1$ and $ C_{1}^{p}(x) = x$. The polynomials
satisfy the following differential equation:
\[
\left\{ (1-x^2)({d\over dx})^2 - (2p+1)x {d\over dx} + N(N+2p)\right \}
C_{N}^{p}(x)= 0.
\]
A change of variable $x=\cos\theta$ puts this in the
following form:
\begin{equation}\label{diffeq}
\left\{ ({d \over d\theta})^2  + 2p {\cos\theta\over \sin\theta }
{d \over d\theta} + N(N+2p)\right \}     C_{N}^{p}(\cos\theta)= 0.
\end{equation}
The polynomials are normalized as:
\[
C_{N}^{p}(1)= {\Gamma(2p+N)\over\Gamma(2p) N!}
= {\mathrm dim}{\H^{(D)}_N}  {D-2 \over 2N+D-2}.
\]
where ${\mathrm dim}{\H^{(D)}_N}$ is given by (\ref{app:dim}).
The polynomials satisfy the following orthogonality condition:
\begin{equation}\label{norm}
        \int_{-1}^{+1} dx (1-x^2)^{p-{1\over 2}}C_{N}^{p}C_{M}^{p}
        =\delta_{N,M} {\pi \Gamma(2p+N)\over
        2^{2p-1} N! (N+p) \Gamma^2(p)}.
\end{equation}

\section{Asymptotics of the Gegenbauer polynomial}
\label{app:asympt}

To get the asymptotics of the Gegenbauer polynomial for large
$N$ we use the differential equation (\ref{diffeq}). It can be
put into a form similar to that of a wave equation by setting
\[
C_N^{(p)}(\cos\theta) = f(\theta) \sin^{-p}\theta.
\]
One gets:
\[
{d^2 f\over d\theta^2} +
f \left[ (N+p)^2 - {p(p-1)\over\sin^2\theta} \right] = 0.
\]
For large $N$ one can neglect the second term in the
square brackets and $p$ as compared to $N$ in the first term.
Thus, the large $N$ asymptotics is given
by
\[
C_N^{(p)}(\cos\theta) \sim {A\over\sin^p\theta}
\sin[(N+p)\theta + \phi],
\]
where $\phi$ is a phase and $A$ is a normalization
factor, both arbitrary at this stage. It
can be constrained by using symmetry properties of
$C_N$. From the expression for the generating functional
one sees that
\[
C_N(-x) = (-1)^N C_N(x).
\]
Thus,
\[
C_N^{(p)}(\cos(\pi-\theta)) = (-1)^N C_N^{(p)} (\cos\theta).
\]
A simple analysis shows that this restricts $\phi$ to be
\[
\phi = {(1-p)\over 2} \pi + \pi k,
\]
where $k$ is an arbitrary integer. Thus, the ambiguity in
$k$ is just the overall sign ambiguity.
The constant $A$ can be determined from the normalization condition
(\ref{norm}). One gets:
\[
{\pi\over 2}\, A^2 = {\pi \Gamma(2p+N)\over
        2^{2p-1} N! (N+p) \Gamma^2(p)},
\]
or
\[
A = \pm {1\over 2^{p-1} \Gamma(p)}
\left[ {\Gamma(2p+N)\over N! (N+p)} \right]^{1/2}.
\]
For large $N$ this behaves as
\[
A \sim \pm {N^{p-1}\over 2^{p-1} \Gamma(p)}.
\]
Using the fact that
\[
C_{2N}^{(p)}(0) = {(-1)^N \over N!}{\Gamma(N+p)
\over \Gamma(p)} \sim {(-1)^N \over \Gamma(p)} N^{p-1},
\]
and  the expression for the derivative of the Gegenbauer polynomial
\[
	{d\over d \theta} C_N^{p}= -2p \sin \theta \, C_{N-1}^{p+1},
\]
we can fix the overall sign to be plus. Thus, finally, we get:
\begin{equation}\label{app:as}
C_N^{(p)}(\cos\theta) \sim  {N^{p-1}\over 2^{p-1} \Gamma(p)}
\,{1\over\sin^p\theta}\,\sin[(N+p)\theta + (1-p)\pi/2].
\end{equation}


\begin{thebibliography}{99}

\bibitem{Penrose} R.\ Penrose, Angular momentum: an approach to
combinatorial space-time; in ``Quantum theory and beyond'',
ed. Ted Bastin, Cambridge University Press, 1971.

R.\ Penrose, Applications of Negative Dimensional Tensors; in ``Combinatorial
Mathematics and its Applications'',
ed. D. J. A. Welsh, Academic Press, 1971.

R.\ Penrose, Combinatorial Quantum Theory and Quantized Directions; in
``Advances
in Twistor Theory'', ed. L. P. Hughston and R. S. Ward,
Pitman Advanced Publishing Program, 1979.

\bibitem{future} L.\ Smolin, The future of spin networks, gr-qc/9702030.

\bibitem{Loop} C.\ Rovelli, Loop quantum gravity,
review written for the electronic journal Living Reviews,
available as gr-qc/9710008.

\bibitem{Roberto} R.\ De Pietri, Canonical ``loop'' quantum gravity and
spin foam models, gr-qc/9903076.

\bibitem{Mike} M.\ Reisenberger, A lefthanded simplicial action for
Euclidean general relativity, {\sl Class.\ Quant.\ Grav.\ } {\bf 14} (1997),
1753-1770.

\bibitem{BC} J.\ Barrett and L.\ Crane, Relativistic spin networks
and quantum gravity, {\sl J.\ Math.\ Phys.\ }{\bf 39} (1998), 3296-3302.

\bibitem{Baez} J.\ Baez, Spin foam models, {\sl Class. Quant. Grav.}
{\bf 15} (1998), 1827-1858.

\bibitem{SF} L.\ Freidel and K.\ Krasnov, Spin foam models and the
classical action principle, {\sl Adv.\ Theor.\ Math.\ Phys.\ }
{\bf 2} (1998), 1221-1285; also available as hep-th/9807092.

\bibitem{FKP} L.\ Freidel, K.\ Krasnov and R.\ Puzio, BF description
of higher-dimensional gravity theories, hep-th/9901069.

\bibitem{BaezSpinNet} J.\ Baez, Spin networks in nonperturbative 
quantum gravity, in {\sl The Interface of Knots and Physics}, 
ed.\ Louis Kauffman, American Mathematical Society, Providence, 
Rhode Island, 1996, pp.\ 167-203; gr-qc/9504036.

\bibitem{VK} N.\ Vilenkin and A.\ Klimyk, Representation of Lie groups
and Special Functions, Vol. 2, Mathematics and its applications,
Kluwer academic publisher, 1993.

\bibitem{Barrett:evaluation} J.\ Barrett, The classical
evaluation of relativistic spin networks, math.QA/9803063.

\bibitem{Barrett:asymptotics} J.\ Barrett, The asymptotics of
an amplitude for the 4-simplex, gr-qc/9809032.

\bibitem{Hos} H.\ Hochstadt,
The functions of mathematical physics,
Dover Publications, 1971.

\end{thebibliography}
\end{document}